\documentclass[aps,prl,reprint,superscriptaddress]{revtex4-1}

\usepackage{graphicx} 
\usepackage{graphics} 
\usepackage{amsfonts} %
\usepackage{color} 

\newcommand{\abs}[1]{\left| #1 \right|} 
\newcommand{\avg}[1]{\left\langle #1 \right\rangle} 

\newcommand{\ket}[1]{\left| #1 \right\rangle} 
\newcommand{\bra}[1]{\left\langle #1 \right|} 
\newcommand{\braket}[2]{\left\langle #1 \vphantom{#2} \right|
 \left. #2 \vphantom{#1} \right\rangle} 

\begin{document}

\title{Amplifying Single-Photon Nonlinearity Using Weak Measurement}

\author{Amir \surname{Feizpour}}
 \email{feizpour@physics.utoronto.ca}
\author{Xingxing \surname{Xing}}
\author{Aephraim M. \surname{Steinberg}}
 \affiliation{Centre for Quantum Information and Quantum Control, and Institute for Optical Sciences, Department
of Physics, University of Toronto, 60 St George Street, Toronto, Ontario,
Canada M5S 1A7}

\begin{abstract}
We show that weak measurement can be used to ``amplify'' optical nonlinearities
at the single-photon level, such that the effect of one properly post-selected
photon on a classical beam may be as large as that of many un-post-selected
photons.  We find that ``weak-value amplification" offers a marked improvement
in the signal-to-noise ratio in the presence of technical noise with long
correlation times.  Unlike previous weak-measurement experiments, our proposed
scheme has no classical equivalent.
\end{abstract}

\maketitle

An interaction between two independent photons could be used to serve as a
``quantum logic gate," enabling the development of optical quantum computers
\cite{PhysRevLett.62.2124,pittman,langford}, as well as opening up an essentially new field of
quantum nonlinear optics \cite{Chiao}.  Typical optical nonlinearities are many
orders of magnitude too weak to create a $\pi$ phase shift as required in
initial proposals, but more recently it was realized that any phase shift large
enough to be measured on a single shot could be leveraged into a quantum logic
gate \cite{Munro}. 
Much recent work has
shown that atomic coherence effects \cite{PhysRevLett.82.4611,
PhysRevLett.84.1419, PhysRevA.68.041801, PhysRevLett.97.063901} and
nonlinearities in microstructured fiber \cite{PhysRevLett.97.023603,
2009NaPho.3.95M} can generate greatly enhanced Kerr nonlinearities.  While even
a very small phase shift can be made larger than the quantum (shot) noise, by
using a sufficiently intense probe, present experiments  are limited by
technical rather than quantum noise and difficult to carry out even with much
averaging. For example, in Ref. \cite{2009NaPho.3.95M}, a phase shift of $10^{-7}$ rad
was measured by averaging over $3\times 10^9$ classical pulses with
single-photon-level intensities. To date, no one has yet been able to observe
the cross-Kerr effect induced by a single propagating photon on a second
optical beam \cite{Kimble}. In this Letter, we show that using weak-value
amplification (WVA) \cite{PhysRevLett.60.1351, 2008Sci.319.787H,
Dixon}, a single photon can be made to ``act like" many
photons, and it is possible to amplify a cross-Kerr phase shift to an observable
value, much larger than the intrinsic magnitude of the single-photon-level
nonlinearity. In so doing, we also demonstrate quantitatively how WVA may
improve the signal-to-noise ratio (SNR) in appropriate regimes, a result of
broad general applicability to quantum metrology.

\begin{figure}[t]
  \centering %
  \includegraphics[width=\columnwidth]{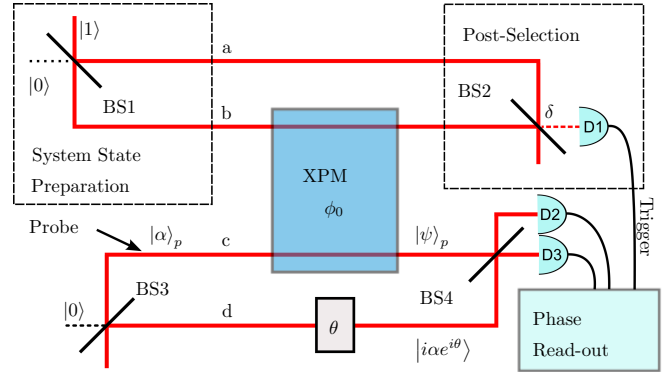}
  \caption{The single-photon ``system'' is prepared in an equal superposition of arms $a$ and $b$ by  the first beam-splitter (BS1).
  After a weak cross-phase-modulation (XPM) interaction with the ``probe'', prepared in a coherent state $\ket{\alpha}_p$, the system is post-selected on a nearly orthogonal state by detecting the single photon in the nearly-dark port, D1.
  The success probability of post-selection depends on the imbalance $\delta$ in the reflection and transmission coefficients of the beam-splitter $BS2$, and the back-action of the probe on the system.
  Using the lower interferometer to read out the phase shift of the probe amounts to a measurement of the system observable $n_b$, the photon number in arm $b$.
  The phase shifter $\theta$ is used to maximize the sensitivity of the measurement.
  \label{fig_setup}}
\end{figure}

Weak measurement is an exciting new approach to understanding quantum systems
from a time-symmetric perspective, obtaining information from both their
preparation and subsequent post-selection \cite{time_in_QM}. In
the past several years, it has been widely studied to address foundational
questions in quantum mechanics \cite{Hardy}, as well as
for its potential application to ultrasensitive measurements
\cite{2008Sci.319.787H, Dixon, 2010Natur.463.890S, Brunner}.
If a quantum system is coupled only weakly to a probe, then very little
information may be obtained from a single measurement, and in compensation,
this measurement disturbs the system by a negligible amount. In such
situations, if the system is prepared in some initial state $|i\rangle$ and
post-selected in some other final state $|f\rangle$, the ``weak value",
$\langle A\rangle_w = \langle f|A|i\rangle / \langle f| i\rangle$, describes
the mean size of the effect an ensemble of such systems would have on a device
designed to measure the observable A. It should be noted that weak values are
not guaranteed to lie within the eigenvalue spectrum of the observable A.
Specifically, if the overlap between the initial and final states is small, the
weak value may be anomalously large. In Aharonov, Albert and Vaidman's famous example, the
spin of an electron may be measured to be 100 \cite{PhysRevLett.60.1351}; in a
mathematically equivalent sense, we show that the effective photon number in
one arm of an interferometer may be found to be 100 even if the entire
interferometer contains only one photon.

Unfortunately, WVA always comes at the cost of reducing the sample size (via post-selection) by just enough to nullify any potential improvement in SNR, at least in the case of statistical noise.
Several recent experiments \cite{2008Sci.319.787H, Dixon}
observed that many real-world measurements are limited by technical noise, which
is not reduced by averaging over more samples, and attempted to show that in
such cases weak measurement can indeed be of practical advantage. 
It still remains unclear exactly when such ``technical'' noise could be overcome by using WVA.
In Refs. \cite{Dixon}, a very specific noise model was assumed,  in which rejection of photons through post-selection did not reduce the ultimate signal strength, an assumption we do not make \footnote{Furthermore, in that the (presumably electronic) technical noise was always averaged over the same time window regardless of how many photons this window contained, detecting more photons could never improve the rejection of the technical noise, regardless of its white spectral characteristics. This is related to the fact that in this essentially ``classical'' post-selection, none of the (electronic) data were ever actually discarded; only the intensity was reduced via the selection, and it was postulated that this occurred without an associated reduction in signal amplitude.}.
Here we find that the SNR can be increased,
roughly to but not beyond the quantum limit, when the noise correlation times
are sufficiently long. Previous weak-measurement demonstrations, instead of
entangling a system with a distinct ``probe," merely used two degrees of
freedom of the same physical photon as the system and probe; this resulted in
experiments which could be equally well understood in the framework of
classical electromagnetism, with no need of the full quantum formalism of weak
measurement. (Some implementations have been carried out with probabilistic
coupling between the system and the probe
\cite{PhysRevLett.94.220405,*PhysRevLett.104.080503}.) Our present proposal
demonstrates that two distinct optical beams may be coupled deterministically,
by using accessible interactions, in such a way that no classical explanation is
possible; that the surprising weak-measurement prediction of a single photon
``acting like" a collection of many photons is rigorously correct; and that the
SNR can be substantially improved by WVA, when the noise possesses long
correlation times (e.g. 1/f noise).

The nonlinear interaction of interest here can be viewed as a measurement in
which a single-photon ``system" is coupled through the cross-Kerr effect to a
classical ``probe" field; see Fig.\ \ref{fig_setup}. The single photon is sent
through a 50-50 beam splitter, thus prepared in the superposition
$\ket{i}\equiv(\ket{b} - \ket{a})/\sqrt{2}$ of modes $a$ and $b$. The single
photon interacts with a probe through a Kerr medium, leading to a cross-phase
shift that we model as $\exp(i\phi_0 \hat{n}_b \hat{n}_c)$, where $\phi_0 \ll
1$ is the cross-phase shift per photon and $\hat{n}_b$ ($\hat{n}_c$) is the
number operator for mode $b$ ($c$). After the interaction with the probe the
system is post-selected to be in a state nearly orthogonal to the initial one,
$\ket{f} = t \ket{b} + r \ket{a} $, by triggering on the detection of a
photon at D1. This port exhibits imperfect destructive interference when the
reflectivity $r$ and transmissivity $t$, which we choose to be real and
positive, are slightly imbalanced. We define a small post-selection parameter,
$\delta \equiv \braket{f}{i} = (t-r)/\sqrt{2} \ll 1$. The weak value of the
photon number in mode $b$ is given by \
\begin{equation}
\avg{\hat{n}_{b}}_w = \frac{\bra{f} \hat{n}_{b} \ket{i}}{\braket{f}{i}}
=\frac{t/\sqrt{2}}{(t-r)/\sqrt{2}} \simeq \frac{(1+ \delta)/2}{\delta} \simeq
\frac{1}{2\delta} \; .
\end{equation}

\noindent This means that whenever the post-selection succeeds (which occurs
with probability $\delta^2$ ignoring the measurement back-action) the weak
value of the photon number in mode $b$ is $1/\delta$ times the strong value,
$1/2$. The post-selection parameter $\delta$ can be very small, leading to a
large weak value for the photon number in the system. Therefore, within the
weak-measurement formalism, the probe will experience a cross-phase shift
equivalent to that of many photons, even though the system never has more than
one photon. In the rest of this Letter, we will show explicitly that such a
scheme does in fact lead to a large phase shift, and quantify the improvement
in the SNR as a function of the characteristics of the technical noise.

The state of the system and probe after coupling is \
\begin{equation}
\ket{\Psi} = \frac{1}{\sqrt{2}} (\ket{b}_s\ket{\alpha e^{i\phi_0}}_p -
\ket{a}_s\ket{\alpha}_p) .
\end{equation}

\noindent For $\phi_0 \ll 1$, the overlap between the two possible final probe
states is $\braket{\alpha}{\alpha e ^{i\phi_0}} \simeq e^{i\abs{\alpha}^2\phi_0
- \abs{\alpha}^2\phi_0^2/2}$. The amplitude of this overlap, $e
^{-\abs{\alpha}^2\phi_0^2/2}$, has to be close to 1 for the interaction to be
weak, which implies  $\abs{\alpha}\phi_0 \ll 1$. The phase of the overlap,
$\abs{\alpha}^2 \phi_0$, describes the average phase-shift imparted to the
system by the probe. This phase does not result in dephasing of the system
state and therefore, in principle, can be compensated by adding a phase-shifter
to the upper interferometer. Without compensation, WVA will occur only when
$\abs{\alpha}^2 \phi_0$ is close to an integer multiple of $2\pi$, where the
overlap between the initial and final states of the system is small. We define
$\epsilon$ to be the difference between $\abs{\alpha}^2 \phi_0$ and the closest
multiple of $2\pi$.

If the system is post-selected to be in state $\ket{f}$, the state of the
probe, $\ket{\psi}_p = ~_s\braket{f}{\Psi}$, collapses to a superposition of
two coherent states, \
\begin{eqnarray}
 \ket{\psi}_p &=& \sqrt{P^{-1}} \frac{1}{2} \left((1+\delta)\ket{\alpha e^{i\phi_0}} - (1-\delta)\ket{\alpha} \right),
 \label{eq_alp}
\end{eqnarray}

\noindent where $P \simeq \abs{\alpha}^2 \phi_0^2 / 4 +\delta^2 + \epsilon^2/4$
is the post-selection probability. The final state of the probe can be most
easily understood by displacing it to the origin in phase space, defining
$\ket{\chi} = D^\dagger(\alpha) \ket{\psi}_p$, where $D(\alpha)$ is the
displacement operator. For $\phi_0 , \abs{\alpha}\phi_0 \ll 1$, one can write \
\begin{equation}
\ket{\chi} \simeq \sqrt{P^{-1}} \left((\delta+i\epsilon/2) \ket{0} +
(i\alpha\phi_0/2)\ket{1}\right), \label{eq_chi}
\end{equation}

\noindent where $\ket{0}$ and $\ket{1}$ are vacuum and single photon number
states respectively. The weak measurement formalism applies if $\delta^2 \gg
(\epsilon^2 + \abs{\alpha}^2 \phi_0^2) / 4$; in particular, as $\epsilon
\rightarrow 0$, one recovers the weak-measurement prediction $\ket{\psi}_p
\simeq \ket{\alpha \exp(i\phi_0 / \delta)}$, a coherent state with a largely
enhanced phase. On the other hand, if $\delta^2 \ll \epsilon^2/4 +
\abs{\alpha}^2 \phi_0^2 / 4$ the post-selection is significantly modified by
the back-action of the probe on the system. It is instructive to look at both
regimes and the transition between them and determine what the maximum possible
enhancement is, taking the back-action into account.

Most of the interesting phenomena can be understood by investigating properties
of $\ket{\chi}$. If $\delta$ or $\epsilon$ is much larger than $\abs{\alpha}
\phi_0$, then the state $\ket{\chi}$ is approximately equal to a weak coherent
state, $ \ket{\chi} \simeq \ket{0} + i \alpha \phi_0 \ket{1}/ (2\delta +
i\epsilon) $. It can be seen that $\delta$ contributes to a shift in the
imaginary quadrature (phase of $\ket{\psi}_p$) and $\epsilon$ contributes to a
shift in the real quadrature (average photon number). On the other hand, if
$\abs{\alpha} \phi_0$ is much larger than the two other terms, the state
$\ket{\chi}$ is approximately a single-photon number state. 

\begin{figure}[t]
  \centering %
  \includegraphics[width=\columnwidth]{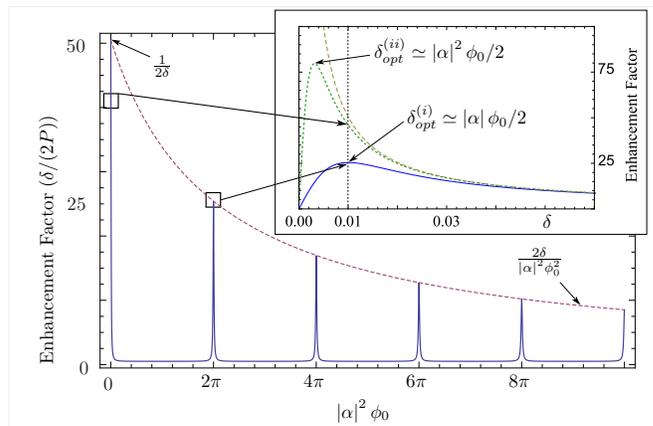}
  \caption{The enhancement factor versus $\abs{\alpha}^2\phi_0$.
  The parameters used are $\phi_0 = 2\pi \times 10^{-5}$ and $\delta = 0.01$.
  The enhancement factor is calculated by using the state of Eq. (\ref{eq_alp}) without any approximations.
  The dashed line shows the enhancement factor if the average phase written by the probe on the system, $\abs{\alpha}^2\phi_0$, is compensated;
  otherwise enhancement occurs whenever $\abs{\alpha}^2\phi_0$ is close to an integer multiple of $2\pi$ (solid curve).
  The inset shows the enhancement factor as a function of post-selection parameter, $\delta$, in two different
  regimes: i) $\abs{\alpha}^2 = 10^5$, in which case the imparted phase on the system by the probe, $\epsilon$, is 0 (solid blue);
  ii) $\abs{\alpha}^2 = 10^2$, where $\epsilon$ is a small non-zero phase (dashed green).
  For large values of $\delta$ the weak-measurement prediction is valid; however as $\delta$ decreases the back-action from the probe plays a more dominant role.
  The dashed line shows the prediction of the weak-measurement formalism.
  \label{fig_enhancement}}
\end{figure}

The average phase
shift can be measured by using the lower interferometer in Fig.\ \ref{fig_setup},
e.g. as the ratio of the difference of the photon numbers at D2 and D3 to the
sum, \
\begin{eqnarray}
\bar{\phi} = \frac{\avg{M_-}_p}{\avg{M_+}_p} \simeq \frac{\delta}{2P} \phi_0,
\end{eqnarray}


\noindent where $M_\pm = \hat{n}_3 \pm \hat{n}_2$. We should compare this value
to the phase shift $\phi_0$ imparted to the probe by a single photon in path
$b$. The phase that one measures after successful post-selection is enhanced by
a factor of $\delta / 2P$. Fig.\ \ref{fig_enhancement} shows this enhancement
factor as a function of post-selection parameter $\delta$ and the average
number of probe photons, $\abs{\alpha}^2$. For sufficiently small back-action,
the weak measurement prediction for the amplification, $1/2\delta$, is correct.
However, as $\delta$ becomes smaller, the amplification grows but so does the
back-action, until at $\delta_{opt} = \sqrt{\abs{\alpha}^2\phi_0^2 +
\epsilon^2} /2$ a maximum amplification value is achieved of $1/4\delta_{opt}$,
half of the weak-measurement value. For small $\epsilon$, the maximum phase
shift is equal to $1/2|\alpha|$, which is one-half the quantum uncertainty of
the probe phase. Thus, the WVA works up to the point where the single-shot
quantum-limited SNR would be on the order of 1. Taking a closer look at the
form of state $\ket{\chi}$, one can see that the large phase shift is caused by
destructive interference due to post-selection; the vacuum term largely cancels
out, enhancing the importance of the single-photon term. Note that the large
overlap of the two possible probe states corresponding to the two states of the
system is essential for this to occur.

The weakness condition $\abs{\alpha} \phi_0 \ll 1$ is often met in experimental
situations, either because of the difficulty of approaching quantum-limited
performance at high intensities or to avoid additional undesired nonlinear
effects. In Ref. \cite{2009NaPho.3.95M}, for instance, a cross phase shift of
$\phi_0 = 10^{-7}$ rad per photon was reported and unwanted nonlinear effects
were observed once the average number of probe photons $|\alpha|^2$ reached
about $10^6$. In this situation both conditions of $\abs{\alpha} \phi_0 \ll 1$
and $\abs{\alpha}^2 \phi_0 \ll 1$ are met and WVA can be used to enhance the SNR.


In practice, phase measurement is subject to both quantum and technical noise.
While the average measured phase is enhanced by a factor of $\delta/2P$, we
expect the uncertainty due to statistical noise to be inversely proportional to
the square root of the sample size, thus scaling as $1/\sqrt{P}$ (recall that $P$
is the probability of successful post-selection).  The overall SNR is hence
multiplied by a factor $\delta/2\sqrt{P}$, which has a maximum value of $1/2$
(the {\it actual} photon number in arm b); in the case of pure quantum noise,
for instance, there is no advantage with post-selection.  In what follows,
using a more general noise model, we study under what type of ``technical''
noise WVA can be beneficial.

Consider a non-post-selected measurement performed over a total time $T$.
Single photons are sent to the upper interferometer at a rate $\Gamma$ and
phase measurement is triggered by the detection of a single photon. We term the
outcome of the $i^{th}$ measurement $\phi^i_m = \bar{\phi} + \eta^i$, where the
zero-mean fluctuating term $\eta^i$ includes  the quantum and technical noise.
The average measured phase shift is $\phi_m = 1/(\Gamma T) \sum_{i=1}^{\Gamma
T} \avg{\phi^i_m} = \bar{\phi}$. The uncertainty in this average value is given
by $(\Delta \phi_m)^2 = 1/(\Gamma T)^2 \sum_{i,j=1}^{\Gamma T} \avg{\eta^i
\eta^j}$. There are two possible extremes to be considered. In the white-noise
limit (noise correlation time $\tau_c$ much shorter than the mean time between
successive measurements, $1/\Gamma$), the correlation function can be modelled
as a delta function: $\avg{\eta^i \eta^j} = \bar{\eta}^2 \delta_{ij}$. In
particular, this holds for quantum (shot) noise. In this limit the noise scales
statistically with the number of measurements, $\Delta \phi_m = \bar{\eta} /
\sqrt{\Gamma T}$. The opposite extreme is that of noise with long-time
correlations, $ \tau_c \gg 1/\Gamma$, in which case $\avg{\eta^i \eta^j} =
\bar{\eta}^2$, and averaging cannot help reduce the uncertainty.

In the post-selected case, the sample size drops from $\Gamma T$ to $P\Gamma
T$, and $\Delta\phi_m$ increases to $\bar{\eta}/\sqrt{P\Gamma T}$ in the
delta-correlated case while it remains constant at $\bar{\eta}$ in the presence
of long-time correlations.  Given the enhancement factor of $\delta/2P$, the
SNR thus scales as $\delta/2\sqrt{P}$ (always $<1$, as
remarked earlier) in the former case but $\delta/2P$ (which may be $\gg 1$) in
the latter case.

\begin{figure}[t]
  \centering %
  \includegraphics[width=\columnwidth]{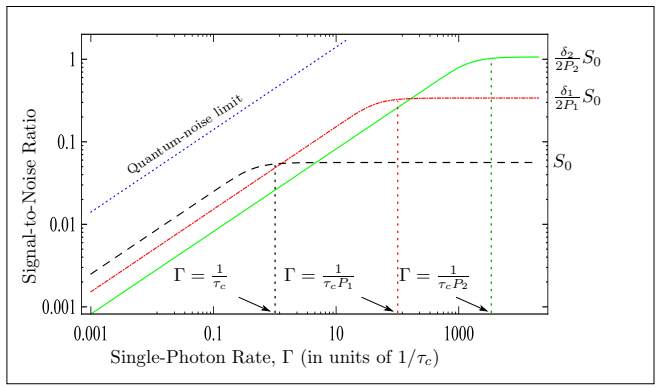}
  \caption{The SNR as a function of the single photon rate $\Gamma$.
  The technical noise is modelled by an exponential correlation function with an
  amplitude, $\bar{\eta}$, 10 times larger than the quantum noise. The dashed
  line shows the non-post-selected SNR for the phase shift due to one photon in mode $b$. The post-selected SNR for $\delta_1 = 0.1$ (weak-measurement regime- dash-dotted red) and $\delta_2 = 0.01$ (the optimum value of measured phase shift- solid green) are also shown;
  the dotted line shows the quantum-limited SNR for comparison.
  The non-post-selected SNR approaches a maximum value, $S_0$, due to low-frequency noise.
   However, for the
  post-selected SNR, we see enhancement by a factor of $\delta/2P$, compared
  to the non-post-selected SNR, $S_0$, for measurements with high enough rate.
  For low rates the enhancement is given by $\delta/2\sqrt{P}$ and therefore the weak measurement results in the best possible post-selected SNR.
  We have taken $T/\tau_c=10^3$, $\phi = 2\pi \times 10^{-5}$, $\abs{\alpha}^2 = 10^5$ and therefore
  $P_1 = 0.01$ and $P_2 = 3\times 10^{-4}$.
  \label{fig_tech_noise}}
\end{figure}

Fig.\ \ref{fig_tech_noise} shows the calculated SNR as a function of single
photon rate, $\Gamma$, where the noise  is modelled with a correlation function
$\avg{\eta^i \eta^j} = \delta_{ij}/2\abs{\alpha}^2 + \bar{\eta}^2
\exp(-\abs{i-j}/\Gamma\tau_c)$ to account for delta-correlated quantum noise
and a technical contribution with correlation time $\tau_c$. The
non-post-selected SNR shows a knee around $\Gamma \tau_c = 1$, separating the
regimes where measurements are not correlated ($\Gamma \tau_c \ll 1$) and
highly correlated ($\Gamma \tau_c \gg 1$).
 The SNR has a statistical scaling, $\sqrt{\Gamma}$, in the former regime and remains constant in the latter.
The graphs for the post-selected cases are qualitatively similar, but the knee
occurs near $P \Gamma \tau_c = 1$, that is, when the noise in the successive
post-selected measurements starts to become correlated. Thus whenever the noise
exhibits correlations over timescales greater than the mean time between
incident photons, the SNR can be improved via post-selection.

We have shown that one post-selected photon may act like many photons, writing
a very large cross-phase-shift on a coherent state, and that this amplification
may greatly improve the SNR for measuring single-photon-level nonlinearities.
Considering presently observable optical nonlinearities, this opens the door to
unambiguous weak measurement experiments, in which two distinct physical
systems could be deterministically coupled, leaving no room for an alternative
classical explanation.  Accounting for the effects of back-action when the
weakness criterion is relaxed, we find that the largest achievable phase shift
per post-selected photon is always of the order of the quantum uncertainty of
the probe phase. More generally, we find that although post-selection cannot
enhance the SNR in the presence of noise with short (or vanishing) correlation
times, particularly shot noise, it can be of great use in the presence of noise
with long correlation times.  Given the prevalence of low-frequency noise (e.g.
$1/f$ noise) in real-world systems, this suggests that WVA may find broad
application in precision measurement.

During the completion of this work, an independent proposal for weakly coupling
photons to atomic ensembles was also posted to the arXiv
\cite{2010arXiv1010.3695S}.

This work was funded by NSERC, CIFAR, and QuantumWorks.  We thank Yakir
Aharonov, Jeff Tollaksen, Sandu Popescu, Lev Vaidman, Chao Zhuang, Alex Hayat
and Greg Dmochowski for many helpful discussions.

\bibliographystyle{apsrev4-1}
\bibliography{refs}

\end{document}